\documentclass[a4paper,fleqn,usenatbib]{mnras}

\usepackage{newtxtext,newtxmath}

\usepackage[T1]{fontenc}
\usepackage{ae,aecompl}

\usepackage{graphicx}	
\usepackage{amsmath}	
\usepackage{amssymb}	
\usepackage{float}
\usepackage{caption}
\usepackage{subfig}

\title[Ionised accretion flows]{Radiation hydrodynamic simulations of massive star formation via gravitationally trapped H{\sc ii} regions -\\
Spherically symmetric ionised accretion flows}

\author[K. Lund et al.]{K. Lund$^{1},$\thanks{E-mail: kblg@st-andrews.ac.uk (KL)}
K. Wood$^{1}$, 
D. Falceta-Gon\c{c}alves$^{1,2}$,
B. Vandenbroucke$^{1}$, 
N. S. Sartorio$^{3}$,
\newauthor  I. A. Bonnell$^{1}$, K. G. Johnston$^{4}$ and E. Keto$^{5}$
\\
$^{1}$SUPA, School of Physics and Astronomy, University of St Andrews, North Haugh, St Andrews KY16 9SS, UK\\
$^{2}$Escola de Artes, Ci\^ encias e Humanidades, Universidade de S\~{a}o Paulo, Rua Arlindo Bettio 1000, CEP 03828-000 S\~{a}o Paulo, Brazil\\
$^{3}$Instituto Nacional de Pesquisas Espaciais - INPE, Divis\~ao de Astrof\'isica, Av. dos Astronautas, 1.758 -Jardim da Granja, \\S\~ao Jos\'e dos Campos - SP, Brazil\\
$^{4}$School of Physics and Astronomy, E.C. Stoner Building, The University of Leeds, Leeds LS2 9JT, UK\\
$^{5}$Harvard-Smithsonian Center for Astrophysics, 160 Garden St, Cambridge, MA 02420, USA\\
}
\date{Accepted XXX. Received YYY; in original form ZZZ}

\pubyear{2018}

\begin{document}
\label{firstpage}
\pagerange{\pageref{firstpage}--\pageref{lastpage}}
\maketitle

\begin{abstract}
This paper investigates the gravitational trapping of H{\sc ii} regions predicted by steady-state analysis using radiation hydrodynamical simulations. We present idealised spherically symmetric radiation hydrodynamical simulations of the early evolution of H{\sc ii} regions including the gravity of the central source. As with analytic steady state solutions of spherically symmetric ionised Bondi accretion flows, we find gravitationally trapped H{\sc ii} regions with accretion through the ionisation front onto the source. We found that, for a constant ionising luminosity, fluctuations in the ionisation front are unstable. This instability only occurs in this spherically symmetric accretion geometry. In the context of massive star formation, the ionising luminosity increases with time as the source accretes mass. The maximum radius of the recurring H{\sc ii} region increases on the accretion timescale until it reaches the sonic radius, where the infall velocity equals the sound speed of the ionised gas, after which it enters a pressure-driven expansion phase. This expansion prevents accretion of gas through the ionisation front, the accretion rate onto the star decreases to zero, and it stops growing from accretion. Because of the time required for any significant change in stellar mass and luminosity through accretion our simulations keep both mass and luminosity constant and follow the evolution from trapped to expanding in a piecewise manner. Implications of this evolution of H{\sc ii} regions include a continuation of accretion of material onto forming stars for a period after the star starts to emit ionising radiation, and an extension of the lifetime of ultracompact H{\sc ii} regions.

\end{abstract}




\section{Introduction}

The classic picture of H{\sc ii} region evolution, as described in textbooks such as \cite{Spitzer1978}, envisages a source of ionising radiation turning on in a uniformly dense medium and ignores the gravity of the source. The H{\sc ii} region initially grows with an ionisation front expanding into neutral gas. This so-called rarefied or R-type phase happens rapidly and the gas structure is relatively unaffected. As the ionised gas is hotter than the neutral gas the next stage sees the H{\sc ii} region grow by pressure-driven expansion, producing a double front structure where a shock front expands into neutral gas ahead of the ionisation front. At late times and large size scales, the assumption that gravity is neglected is valid and this model is very successful in explaining the dynamics and structures of old and large H{\sc ii} regions that are studied at optical and infrared wavelengths. However during early H{\sc ii} region evolution the neglect of gravity presents a problem for the formation of massive stars because it assumes feedback from ionising radiation will halt accretion as soon as the star turns on, thereby limiting the mass that can be attained.

In a series of papers, \cite{Keto2002,Keto2002b,Keto2003} expanded on work presented by \cite{Mestel1954} and examined the evolution of H{\sc ii} regions at size scales where the gravity of the central source dominates the gas dynamics. He first studied steady state solutions for two-temperature spherically symmetric Bondi accretion \citep{Bondi1952} whereby a source of ionising radiation creates a hot ($T_i\approx 10^4$~K) H{\sc ii} region within cold ($T_n\approx 100$~K) neutral inflowing gas. When the H{\sc ii} region is inside approximately the ionised sonic point, where the flow velocity equals the local sound speed, it is unable to expand and remains gravitationally trapped. Accretion occurs from the infalling cold neutral gas and continues inwards towards the star through the H{\sc ii} region boundary where the gas forms an ionised accretion flow. As the star accretes mass its luminosity increases and the H{\sc ii} region grows in size, not due to pressure driven expansion, but because of the increasing ionising luminosity. Eventually the H{\sc ii} region will grow to surpass the critical radius close to the sonic point, gravity of the central star will no longer dominate, and the H{\sc ii} region will begin a pressure-driven expansion. At this stage accretion through the ionisation front can no longer occur. Some of the already ionised gas will move outward along with the expanding ionisation front, but the innermost gas continues to accrete onto the star, and as the H{\sc ii} region is drained accretion halts. This marks the maximum mass the star can accrete in a spherically symmetric system and beyond this stage the H{\sc ii} region continues its pressure-driven expansion. 

In subsequent papers \citep{Keto2006,Keto2007} this work was extended to consider H{\sc ii} region evolution within rotationally flattened structures. In this two-dimensional picture the H{\sc ii} region grows more rapidly into the lower density regions, creating a bipolar outflow. Accretion of ionised material continues through the dense midplane until the ionising luminosity increases the size of the H{\sc ii} region beyond the critical radius for gravitational trapping in the midplane, thereby shutting off accretion similar to the spherical case. In some scenarios, however, the midplane regions are so dense that they are shielded from radiation. Consequently, accretion onto the star through these regions would not be halted by ionization feedback, but by some other process\citep{Nakano1989,Yorke2002,Krumholz2009,Kuiper2010,Harries2017,Kuiper2018}.

This picture of massive star formation through gravitationally trapped H{\sc ii} regions is appealing in that it overcomes the mass limits set by ionisation feedback in the classic picture where gravity is ignored. It also means that the timescale for the growth of the H{\sc ii} region initially depends on the accretion timescale and is not solely due to the rate of the pressure-driven expansion. This provides a potential solution to the lifetime problem for ultracompact (UC)H{\sc ii} regions; the number of UCH{\sc ii} regions found in the \cite{Wood1989} survey suggests they have a longer lifetime than predicted by pressure-driven expansion of an UCH{\sc ii} region into a diffuse H{\sc ii} region. 

This theoretical framework can also explain observations of massive star forming regions that show infall velocities in molecular gas at large sizescales that smoothly match on to infall velocities in ionised gas at smaller sizescales \citep{Keto2002,Sollins2005}. Observations showing both inflow and outflow \citep{Sollins2005b,Klaassen2007,Klaassen2018} are explained with the picture of H{\sc ii} region evolution in axisymmetric rotationally flattened structures. 

The analysis presented by \cite{Keto2002b,Keto2003} and \cite{Mestel1954} modelled the H{\sc ii} region evolution as a sequence of steady state solutions with increasing ionising luminosity. We extend this work by running simulations of the same process to provide a consistent radiation hydrodynamic evolution. In this paper we present numerical simulations from our radiation hydrodynamics code of spherically symmetric ionised accretion flows. Our code reproduces the main features described in the works of Mestel and Keto. We find that gravitationally trapped H{\sc ii} regions naturally arise within the ionised sonic point, but are unstable in spherical geometries.

Future papers will study the spherically symmetric instability in more detail and move on to two-dimensional axisymmetric and three-dimensional simulations. Section~2 describes our radiation hydrodynamics code, Section~3 presents results for spherical ionised accretion flows, and we present our conclusions in Section~4.

\section{Radiation hydrodynamics code}

Our code is described in detail in an upcoming paper, but the main features are briefly outlined here. In order to simulate the plasma dynamics as it is impacted by the effect of ionisation, we have coupled a magnetohydrodynamical (MHD) code, modified from a publicly available GODUNOV code\footnote{https://bitbucket.org/amunteam/godunov-code} \citep{Kowal2011,2015MNRAS.446..973F,2015ApJ...808...65F,2017ApJ...838...91K, 2017MNRAS.465.4866S}, with a time independent Monte Carlo radiation transfer (MCRT) code for photoionisation. Even though the code is able to deal with magnetic fields, for the purposes of this work, these have been neglected. Real collapsing cores in the interstellar medium are indeed magnetised, however the main goal of this work is the direct comparison of our numerical models to an unmagnetised analytical framework. Also, not only are full radiative MHD models computationally expensive compared to an HD approximation, MHD waves would be allowed to grow, propagate and non-linearly interact with other perturbations of the fluid. This would make the comparison to the analytical model irrelevant.

The code evolves compressible fluid dynamics in a 3D Cartesian grid by solving the set of hydrodynamical equations, solved in the conservative form as: 
\begin{equation}
\partial_t\mbox{\bf U} + \nabla \cdot\mbox{\bf F}(\mbox{\bf U}) = f(\mbox{\bf U}),
\label{eq:genform}
\end{equation}
where ${\bf U}$ is the vector of conserved variables:
\begin{equation}
{\bf U} = \Bigl[  \rho, \rho{\bf v}, \left(\frac{1}{\gamma-1} p + \frac{1}{2}\rho v^2 \right)\Bigr]^T,
\end{equation}
${\bf F}$ is the flux tensor:
\begin{equation}
{\bf F} = \Bigl[ \rho {\bf v}, \rho {\bf v}{\bf v}+ p{\bf I}, \left(\frac{\gamma}{\gamma -1}p + \frac{1}{2}\rho v^2 \right){\bf v}\Bigr]^{T},
\label{eq:flux}
\end{equation}
and $f$ corresponds to source terms for the given conserved variable $U$. $\rho$ represents the gas mass density, ${\bf I}$ the identity matrix, ${\bf v}$ the fluid velocity, $p$ the thermal pressure and $\gamma$ the adiabatic polytropic index. We use the equation of state $p \propto \rho^{\gamma}$ with $\gamma = 1.001$ to mimic an isothermal equation of state, as the cooling time in the scenario we are considering is much shorter than the hydrodynamical timescale.  

The spatial reconstruction is obtained by means of a $3^{rd}$ order monotonicity-preserving method \citep{he11}, with flux piecewise discontinuity being solved approximately by the HLLC Riemann solver \citep{mignone06}. The time evolution is done by means of the $3^{rd}$ order four-stage explicit Strong Stability Preserving Runge-Kutta method \citep{ruuth06}, with a Courant-Friedrichs-Lewy (CFL) of 0.2. 

In the GODUNOV scheme, the hydrodynamical equations are first solved in conserved form, and any external source term (e.g. gravity) is calculated as an update to the physical conserved variables (mass, momentum and energy) afterwards. The gravity term has been included as a point source mass at the grid centre. For the sake of simplicity, the gravitational effect of the interstellar gas on the central source is not being considered in this work, therefore the point source is assumed to be fixed. Photoionisation is also treated in our numerical scheme as a source term for the energy equation, being obtained from the MCRT module. The energy density transferred to the gas as a MC photon packet is absorbed is $\delta \epsilon \simeq 2 \times \frac{3}{2} n k (T_{\rm MC} - T_{n})$, where $n$ is the number density, $k$ is the Boltzmann constant, $T_{\rm MC}$ represents the temperature of equilibrium of the ionised gas (see equation \ref{eq:temp} below) and $T_{n}$ represents the neutral gas temperature. The factor of 2 multiplying the thermal energy difference corresponds to the assumption that ionisation generates ions and electrons in a number twice that of neutral gas.

The principle of MCRT is modelling photon propagation using probability distributions. For the specific workings of the Monte Carlo photoionisation code we are using see \cite{Wood1999,Wood2000,Wood2004}. For each timestep of the hydrodynamical code, the 3D density grid is exported to the MCRT module. The MCRT module is therefore considered time-independent, and equilibrium reached quasi-instantaneously compared to the dynamical timescale of the system. This assumption is adequate given that the typical radiative timescales, namely the light crossing-time $\tau_{\rm light} \sim L c^{-1}$ and the recombination time $\tau_{\rm recomb} \sim (n_e \ \alpha_{\rm A})^{-1}$, are much shorter than the hydrodynamical timescale $\Delta t \sim L \ {\rm max}(u)^{-1}$.\footnote{Here $L$ represents the scale of the system, $c$ the speed of light, $n_e$ the electron density, $\alpha_{\rm A}$ the recombination coefficient to all levels, and max$(u)$ the maximum local speed detected over all cells of the simulated cube.}

For simplicity the material in the grid is assumed to be 100 \% hydrogen, with no heavier elements and no dust. Our aim is to determine the ionisation structure of hydrogen, and assuming pure hydrogen is sufficient for that purpose. Including heavy elements is important for cooling rates, but their low abundances and hence small opacities relative to hydrogen do not influence the radiation transport and the resulting ionisation structure of hydrogen \citep{Wood2004}. We let the fixed gravitational point source at the centre of the grid also be an ionising point source with a specific luminosity. This ionising source emits packets of photons isotropically, which are tracked through a random walk of photoionisations and photon re-emissions until the packet either escapes the grid or is re-emitted as non-ionising photons. We are assuming instantaneous photoionisation equilibrium, so there is no time dependence. We do not consider a full spectrum for the photon frequencies, but apply a two-frequency approximation for the MCRT photoionisation code (see detailed description and benchmark tests in \cite{Wood2000}). Stellar photons, those originating from the source, are given an energy of 17.9 eV, representing an average frequency for ionising photons originating from a 40000 K star (see Appendix of \cite{Wood2000}). Diffuse photons, meaning the re-emitted ionising photons, are given an energy of 13.6 eV, as they have a spectrum which is strongly peaked around that value. We record the distances travelled by each photon packet within the grid cells they pass through, along with whether that distance was travelled by stellar or diffuse photons.

After tracking all photon packets we balance the number of photoionisations and recombinations per unit of time to obtain the ionisation fraction in each grid cell: 
\begin{equation}
n_{H^0}\int_{\nu_o}^{\infty}\frac{4\pi J_{\nu}}{h\nu}\sigma_{\nu}d\nu=\alpha_An_en_p.
\label{eq:equilibrium}
\end{equation}
$\alpha_A$ is the recombination coefficient to all levels (e.g. $\alpha_A\approx 5.25\times10^{-13}\rm{cm}^3\rm{s}^{-1}$ for hydrogen at T = 8000 K, based on Table 2.1 in \cite{Osterbrock2006}), $h$ is the Planck constant, $J_\nu$ is the mean intensity at frequency $\nu$, and $h\nu_o =$ 13.6 eV. $n_{H^0}$ is the number density of neutral hydrogen, $n_e$ is the number density of electrons and $n_p$ is the number density of protons which we can rewrite as the number density of ionised hydrogen, $n_{H^+}$. Assuming pure hydrogen in ionisation equilibrium the number of electrons equals the number of protons, hence $n_pn_e=n_{H^+}^2$. Using path length estimators in the MCRT code \citep{Lucy1999} the integral can be rewritten for each cell as:
\begin{equation}
I=\int_{\nu_o}^{\infty}\frac{4\pi J_{\nu}}{h\nu}\sigma_{\nu}d\nu=\frac{Q}{NV}\sum l\sigma_\nu,
\label{eq:integral}
\end{equation}
where $Q$ is the ionising luminosity, N is the number of photon packets and $V$ is the volume of the cell. From our tracking of the photon packets we can calculate $\Sigma l\sigma_\nu$; the sum of all the path lengths travelled through that specific grid cell times the cross section corresponding to the type of photons (stellar or diffuse) responsible for each path length. By splitting the total number density of hydrogen into a neutral and an ionised component:  
\begin{equation}
n_H=n_{H^0}+n_{H^+},
\label{eq:nH}
\end{equation}
the equilibrium equation can be rewritten as a quadratic equation:
\begin{equation}
\alpha_A n_{H^+}^2 + In_{H^+} - In_{H} = 0.
\label{eq:nHequilibrium}
\end{equation}
At the end of each MCRT iteration we solve Equation \ref{eq:nHequilibrium} in each cell to get the neutral gas fraction, $n_{\rm{frac}}={n_{H^0}}/{n_{H}}$. The ionisation fraction, and hence opacity, of each cell is updated to their new values. The code runs through ten iterations before we accept the current ionisation structure and opacity values.

In order to get reliable results the code needs to run through a substantial number of photon packets, we use 100000 packets for the first 7 iterations and ten times more for the last 3. Considering the photon density will decrease at larger distances from the source, the larger the simulation box size, the more photon packets are needed to get good statistics for the Monte Carlo estimators for the ionisation rate in the outer parts, similar to ray tracing techniques. Since we are studying spherically symmetric accretion we have introduced a radial averaging scheme which uses symmetry to get better signal to noise without having to increase the number of photon packets. This results in spherically symmetric idealised 3D MCRT simulations. We observe that increasing the number of photon packets by a factor of ten does not change the calculated ionisation structure noticeably, but increases the computational time. Consequently we assume ten iterations using our original number of photon packets combined with the radial averaging scheme is sufficient.

In reality there are a number of heating and cooling processes governing the temperature of the gas, but the net result is a temperature of approximately 8000 K for photoionised solar metallicity gas, see for example \cite{Wood2004}. Although we only need to consider Hydrogen opacity to determine the ionisation structure, setting the temperature of ionised gas to be 8000 K approximates the existence of other elements that provide cooling to give this temperature for photoionised gas. By assigning fixed temperatures for neutral and ionised gas, $T_n$ and $T_i$ respectively, we can determine the temperature for any given ionisation fraction using the common approximation \citep{Haworth2012}
\begin{equation}
T_{\rm MC}=T_n+(T_i-T_n)(1-n_{\rm{frac}}).
\label{eq:temp}
\end{equation}
Once the ionised and neutral gas temperatures are given by the MCRT module, the pressure tensor in the hydrodynamics code is updated. 

The code described above has already been successfully tested and benchmarked for the standard Spitzer solution of spherical expansion of an H{\sc ii} region in a uniform medium (Falceta-Gon\c calves, in prep). This benchmark test \citep{Bisbas2015} shows our code reproduces the slow D-type expansion of an H{\sc ii} region. In Section~3.3 below, we show that when the ionisation radius is held fixed our code also reproduces the analytic steady state solution for a fast two temperature R-type accretion flow.

Throughout our paper, we adopt the usual definitions for D-type and R-type ionisation fronts \citep{Osterbrock2006} where the relative velocity, $v_r$, between the ionisation front and neutral gas is defined by two critical conditions:
\begin{equation}
v_r\leq c_i-\sqrt{c_i^2-c_n^2} = v_D \approx\frac{c_n^2}{2c_i}
\label{eq:Dtype}
\end{equation}
and 
\begin{equation}
v_r\geq c_i+\sqrt{c_i^2-c_n^2} = v_R \approx2c_i, 
\label{eq:Rtype}
\end{equation}
where $c_i$ and $c_n$ are the sound speeds in the ionised and neutral gas. Velocities lower than $v_D$ result in a subsonic D-type front, and velocities higher than $v_R$ results in a supersonic R-type front. 

\section{Rad-hydro simulations of spherical ionised accretion flows}

The purpose of this paper is to investigate the gravitational trapping of H{\sc ii} regions predicted by steady-state analysis using radiation hydrodynamical simulations. Starting with a central star where gravity causes a spherically symmetric accretion flow we are interested in the gas dynamics when we turn on radiation at size scales where the infall velocities of the accreting gas exceed the R-critical condition (Equation \ref{eq:Rtype}).

We use a non-adaptive Cartesian grid to simulate our region of interest, we are not concerned with what happens to the gas outside our simulation box, nor the details of how material gets onto the central star. In the following simulations we will not update the mass and luminosity of the star during accretion. We are dealing with an accretion rate of $\sim 4\times10^{-7}$ $\rm{M}_{\odot}\rm{yr}^{-1}$, which implies a timescale of megayears to accrete 1 $\rm{M}_{\odot}$ of material onto the star. Considering our simulations run over tens to hundreds of years any significant change in mass and luminosity will occur over timescales much greater than those investigated here.  

\subsection{Steady state Bondi accretion}

\begin{center}
\begin{table*}
  \begin{tabular}{ | c | c | c | c |c |c |c |c |}
    \hline
    $M_{\star}$ ($M_{\odot}$) & $\rho _B$ ($\rm{cm}^{-3}$) & $T_n$ (K) &$T_i$ (K) & Inner mask (AU) & Outer mask (AU) & Grid cells\\ \hline
    17.87 & 30000 & 500 & 8000 & 10.4 & 38.0 & 128\\ \hline
     \end{tabular}
  \caption{Basic simulation properties. Columns show, respectively, stellar mass, density at the neutral Bondi radius, neutral gas temperature, ionised gas temperature, radius of inner mask, radius of outer mask and number of grid cells in each direction. }
  \label{table:params}
   \end{table*}
\end{center} 

\begin{figure*}
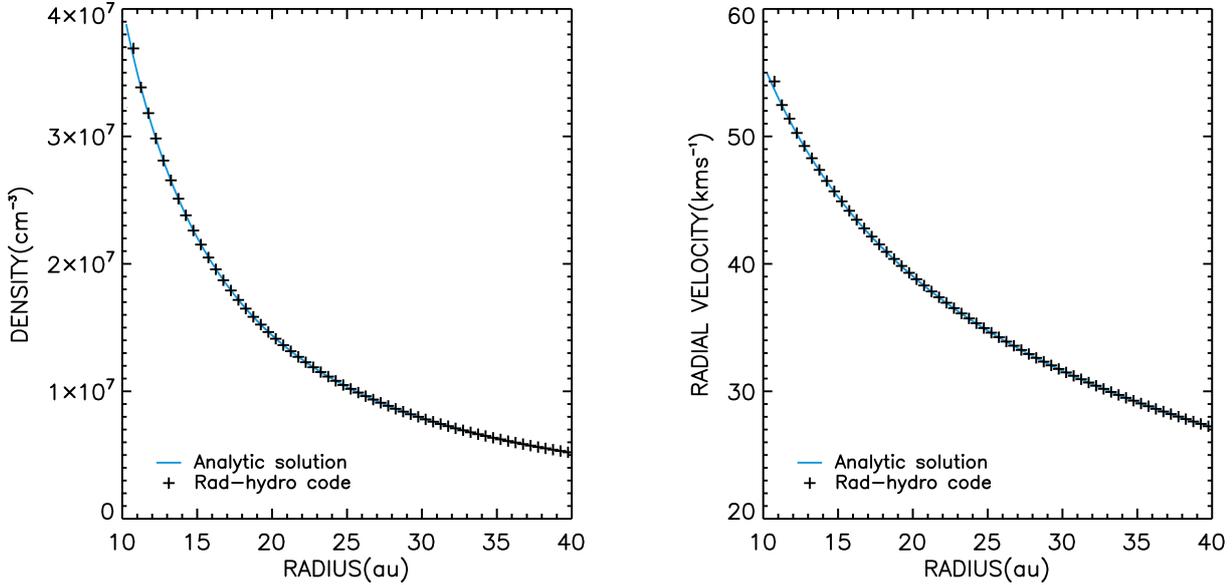

  \subfloat
  {\includegraphics[width=\columnwidth]{bondidensity.pdf}}
  \subfloat
  {\includegraphics[width=\columnwidth]{bondivelocity.pdf}}
	\caption{Radially averaged steady-state density (\textit{left}) and inflow velocity (\textit{right}) from the rad-hydro simulation of material accreting onto a central source compared to the analytic Bondi accretion solution \citep{Vandenbroucke2019}.}
	\label{fig:bondi}
\end{figure*}

We include the gravity of a central star in the code by adding source terms to the momentum and energy hydrodynamical equations. When we consider the gravitational pull of the central star to be the only force acting the result is known as Bondi accretion: spherically symmetric inflow with a constant accretion rate \citep{Bondi1952}. Before switching on radiation we want to make sure our code can reproduce the behaviour associated with Bondi accretion. We compare the results from our simulations with analytical expressions for the expected density and velocity as a function of radius, shown as the blue lines in Figure \ref{fig:bondi}. For a derivation of these expressions from the constant Bondi accretion rate see \cite{Vandenbroucke2019}.
 
Due to the gravitational force there will be a progressively steeper increase in density and velocity closer to the source, with a singularity at the origin. We resign ourselves to not being able to resolve the physics close to the source and encase the region in a mask, inside which we specify a sink of constant density and in general no change of any variables as the simulation runs. Effectively any material that would have accreted onto the star disappears once it passes through the mask.

As an outer boundary condition we set an external sphere with constant physical properties. The outer mask is a way of effectively transforming the cubed computational domain into a sphere, with less spurious numerical fluctuations. In every simulation presented in this paper the region outside the sphere is set to follow the analytic Bondi accretion profile and provides a constant inflow of material, consistent with the pull from the central mass. To set up the initial accretion we define the central mass of the star, $M_{\star}$, the neutral sound speed, $c_n$, and the density at the neutral Bondi radius\footnote{The neutral Bondi radius is where the infall velocity equals the neutral sound speed.}, $\rho _B$.

When picking the initial parameters of the simulation we want to ensure physically plausible values within a size scale where we can observe radial velocities exceeding the R-critical value, in this paper we define a positive radial velocity as pointing towards the central source. We use the term R-critical radius for the position where the infall velocity equals the R-critical velocity. At the centre of the grid we place an 17.87 $\rm{M}_{\odot}$ star. The simulation region must be close enough to this star that the infall velocity will be high enough to observe R-type behaviour once we turn on ionisation. We simulate ${x,y,z} \in [-40,40]$ AU, each direction being divided into 128 cells. The inner and outer masks have radii of 10.4 and 38.0 AU respectively. The density at the neutral Bondi radius is $30000\, \rm{cm}^{-3}$, and all the gas is neutral with a temperature of 500 K\footnote{The choice of a somewhat high temperature is because the numerical simulations runtime scales upward as the temperature difference between ionised and neutral parts increases, and we wanted the simulations to be concluded in the computational time available.}, which corresponds to a sound speed of $\sim 2.0\,\rm{km\,s}^{-1}$. Once the gas is fully ionised the temperature increases to 8000K, and the sound speed is approximately $11.5\,\rm{km\,s}^{-1}$, which results in an R-critical velocity of $v_R \approx 23.0\,\rm{km\,s}^{-1}$. The parameters are summarised in Table \ref{table:params} and are also used for the simulations in sections ~3.2 and ~3.3. 

For the initial setup of the Bondi accretion simulation every parameter in the region of interest between the inner and outer mask are kept constant, the density is left as the value at the outer mask boundary, and the velocity is zero. The system is then allowed to evolve in time, with the gravity of the central star being the only external force source present. The code reached a steady state which, in Figure \ref{fig:bondi}, is shown to match the Bondi profiles. The densities and velocities plotted are obtained by radially averaging throughout the simulation box. The entire simulation region has infall velocities above $v_R$, which is the condition placed upon our chosen initial parameters. Note the steady state Bondi profile, which we have been able to reproduce here, will be the starting point for all subsequent simulations. 

\subsection{Accretion onto a source with constant ionising luminosity}

Starting from the stable Bondi accretion structure described in the previous section we turn on radiation of a fixed luminosity from the central star. We pick an ionising luminosity of $Q = 2.5\times 10^{46}\,\rm{s}^{-1}$, chosen such that the star initially ionises material out to a radius of $\sim$ 32 AU. Notice that, because of the use of an internal mask, the value of $Q$ in the simulations corresponds to the remaining ionizing photon luminosity at the mask radius, and not the stellar luminosity itself. Considering the complete ionization of the gas within the inner mask, we find $Q$ corresponds to $\sim 3\%$ of the central source ionizing luminosity, in agreement with a $\sim 18$M$_{\odot}$ star \citep{Vacca1996}.  

  \begin{figure}
\includegraphics[width=\columnwidth]{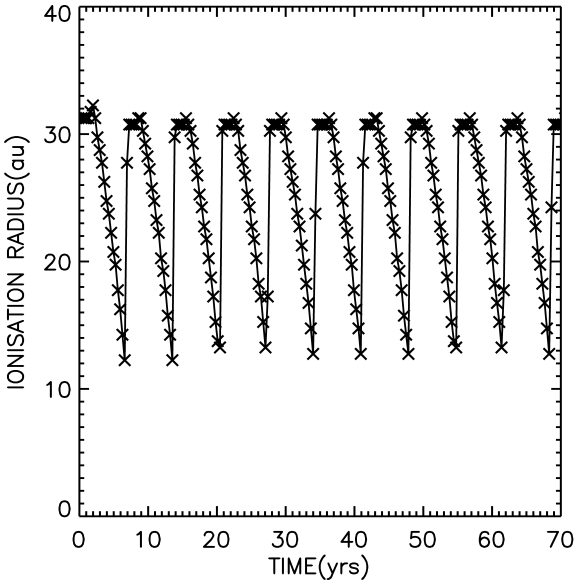}
	\caption{Ionisation front radius as a function of time during the simulation of an 17.87 $\rm{M}_{\odot}$ central star with constant ionising luminosity. Zero years marks the moment radiation is switched on.}
	\label{fig:HIIshrink}
\end{figure}  

  \begin{figure*}
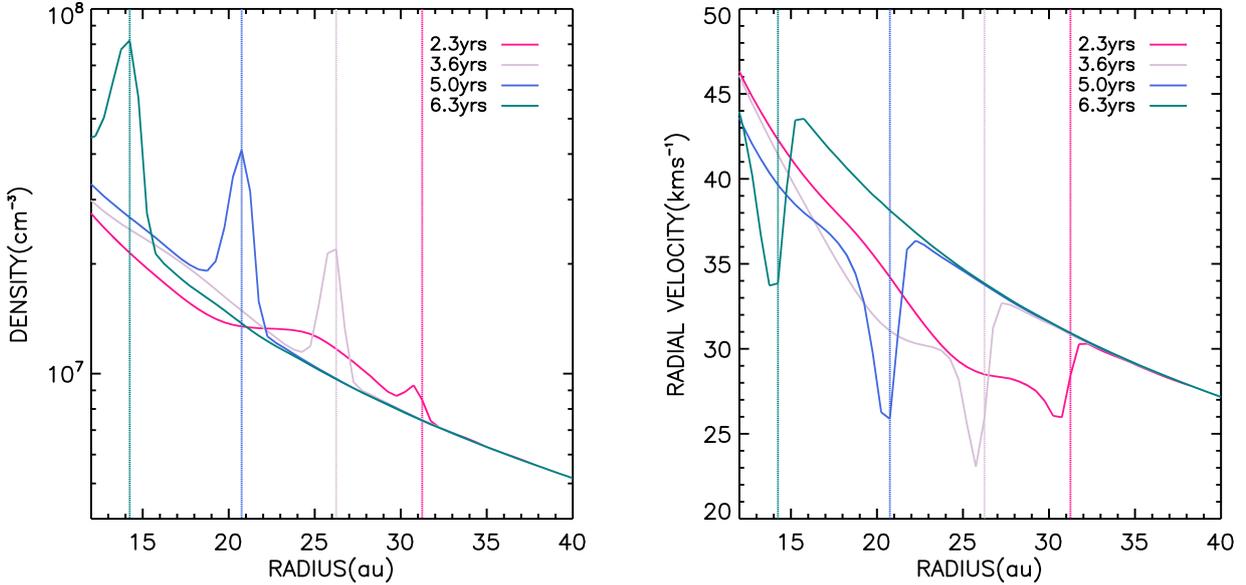

  \subfloat
{	\includegraphics[width=\columnwidth]{shrinkdensity.pdf}}
\subfloat
{	\includegraphics[width=\columnwidth]{shrinkvelocity.pdf}}
	\caption{Radially averaged density (\textit{left}) and inflow velocity (\textit{right}) at four different times during the simulation of an 17.87 $\rm{M}_{\odot}$ central star with constant ionising luminosity. The vertical lines represent the position of the ionisation front. Zero years marks the moment radiation is switched on.}
	\label{fig:shrink}
\end{figure*}

Based on knowing the ionisation fraction in each grid cell the ionisation radius is determined by radially averaging the neutral gas fractions, using bins spanning approximately half a cell width, then looping through these bins in radial order until the neutral fraction equals or exceeds 0.5. The centre of the corresponding bin is defined to be the ionisation radius. Figure \ref{fig:HIIshrink} shows this radius of the ionised region as a function of time. When we turn on the radiation the star ionises a region out to $\sim$ 32 AU, but then the ionisation front proceeds to shrink until it disappears within the inner mask and then moves back out. This process recurs on a 7 year timescale. 

Figure \ref{fig:shrink} shows four snapshots of the density and velocity profile of the system at different times during the first collapse of the H{\sc ii} region. The different colours correspond to different times and the vertical lines mark the ionisation radius. There is a growing density peak around the ionisation radius which moves inward as the H{\sc ii} region shrinks. We can explain the appearance of an initial density perturbation by considering the jump conditions across the ionisation front. Assuming $v_n \geq v_R \gg c_n$, it can be shown that \citep{Osterbrock2006}
\begin{equation}
\frac{\rho_i}{\rho_n}\approx 1+\frac{c_i^2}{v_n^2}. 
\label{eq:Rdensity}
\end{equation}
Consequently there will be a density increase in the ionised gas, and a corresponding decrease in velocity.

The increase in density leads to an increased recombination rate, so the fixed ionising luminosity cannot ionise to the same radius at the next time step of the radiation hydrodynamics simulation. Material keeps piling up at the front and the density increases because the inflow velocity of material is greater outside the ionised region than inside. As the density peak grows the ionisation radius must shrink further, resulting in the runaway behaviour shown in Figure \ref{fig:shrink}.

When the ionisation front and the density enhancement reaches the inner mask we assume it accretes onto the central star. As the density peak disappears inside the mask the amount of material inside the ionised region decreases drastically and in response the ionisation front expands. The region outside the ionisation front has returned to a neutral Bondi profile, meaning the problem is essentially reset and the process repeats. For the parameters of this simulation, the H{\sc ii} region oscillates from just outside the inner mask to a maximum radius $R_{\rm{max}}\approx$ 31 AU over a regular period of $\sim$ 7 yrs. 

\cite{Peters2010} also found flickering in their H{\sc ii} simulations. Whilst we are considering spherically symmetric accretion onto a single star, \cite{Peters2010} ran a 3D simulation on a much larger size scale (several parsec) of the collapse of a rotating molecular cloud. In our case the contraction of the H{\sc ii} region is triggered by jump conditions at the R-type ionisation front leading to a density peak, in contrast \cite{Peters2010} flickering was due to large scale variations in the accretion flow.

According to \cite{Mestel1954} and \cite{Keto2002b,Keto2003}, if the initial H{\sc ii} region starts out close enough to the star to be in the R-type regime it should be possible to prevent expansion, and keep accretion going with material moving through the ionised region. Our simulation of this scenario continuously accretes material onto the star and successfully traps the H{\sc ii} region within a radius $R_{\rm{max}}$, but it is not a steady state solution as we observe recurring expansion and contraction. 

Further investigation into the properties of this instability is being carried out analytically, with a 3D code, and a 1D code capable of much higher time and spatial resolution, see \cite{Vandenbroucke2019}. They find the 7 year timescale observed roughly corresponds to the local free-fall time at the ionisation front radius, as the instability is caused by a small increase in density that is accreted onto the central mask within this timescale and then reseeded. Furthermore the recurring expansion and contraction of the H{\sc ii} region seems to be a result of the idealised spherically symmetric geometry caused by the radial averaging of the photoionisation, in fully 3D simulations the same behaviour is not observed.    

\subsection{Steady state two-temperature accretion}

In the previous section we showed our code produced a gravitationally trapped, but oscillating, H{\sc ii} region. Next we aim to produce a steady state two-temperature accretion solution by forcing the ionisation radius to remain constant in time. 

Again we start from a stable neutral Bondi accretion flow, but instead of switching on a constant ionising luminosity, $Q$, we set the ionising luminosity, $Q_{R_i}$, before each call to the radiation transfer scheme to be the value that will fully ionise gas out to a prescribed radius, $R_i = 24.75$ AU. This is determined by balancing photoionisation with radiative recombination (adopting on-the-spot approximation) inside the radius $R_i$:  
\begin{equation}
Q_{R_i} = \Sigma n^2\alpha_B \Delta V,
\end{equation}
where the sum is over all cells with a radius $r \le R_i$, $n$ is the number density, $\alpha_B = 2.5\times 10^{-13}\,{\rm cm}^3\,{\rm s}^{-1}$ is the Case B recombination coefficient for hydrogen at 8000~K, and $\Delta V$ is the volume of a cell. 

\begin{figure*}
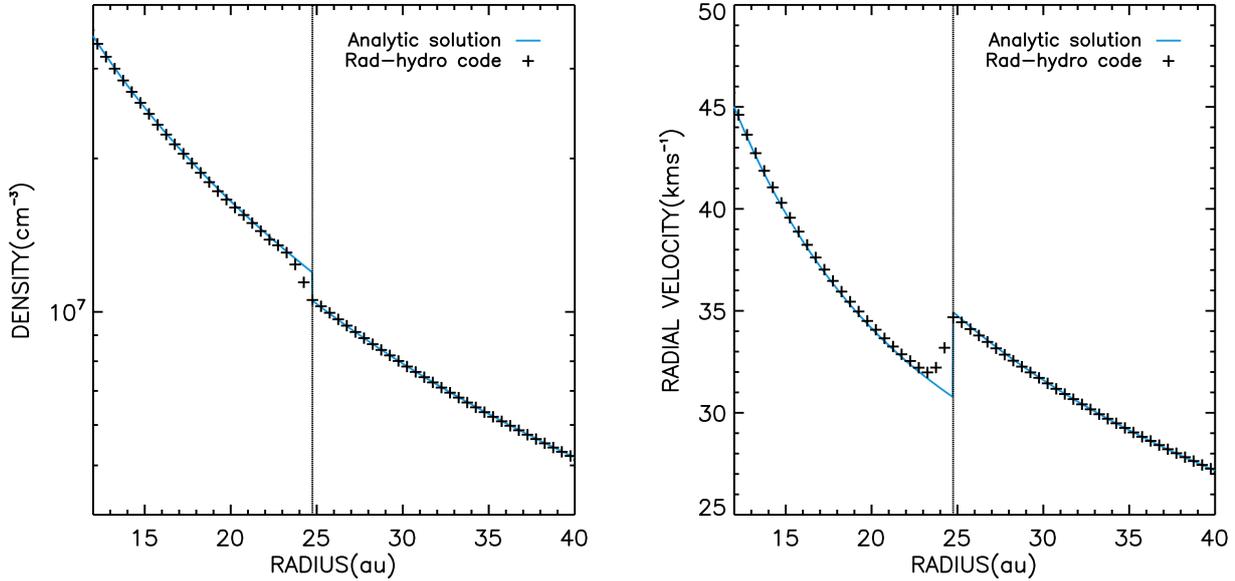

  \subfloat
  {\includegraphics[width=\columnwidth]{2tempdensitybondi.pdf}}
  \subfloat
  {\includegraphics[width=\columnwidth]{2tempvelocitybondi.pdf}}
	\caption{Radially averaged steady state two-temperature density (\textit{left}) and inflow velocity (\textit{right}) from our rad-hydro simulation compared to the analytic solution \citep{Vandenbroucke2019}. The simulation forces an H{\sc ii} region out to the fixed radius $R_i = 24.75$ AU, the vertical line represents the position of the ionisation front.}
	\label{fig:2temp}
\end{figure*}

\begin{figure}
\includegraphics[width=\columnwidth]{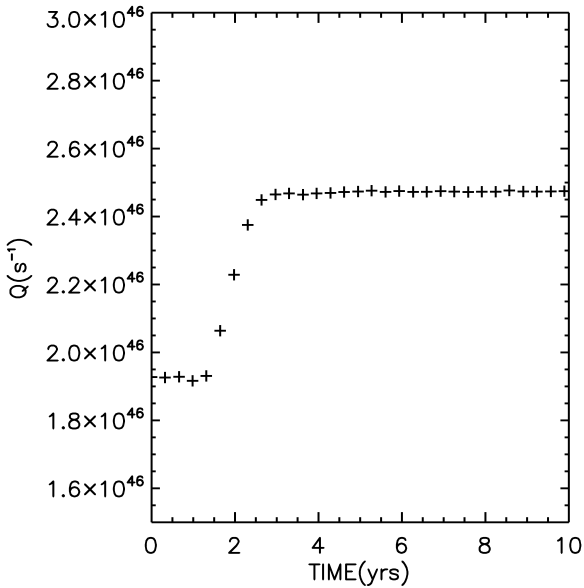}
	\caption{Ionising luminosity, $Q_{R_i}$, required to maintain an H{\sc ii} region out to the fixed radius $R_i = 24.75$ AU as a function of time. Zero years marks the moment radiation is switched on.}
	\label{fig:Qdevelopment}
\end{figure}

The size of the ionised region is kept constant and as a consequence, instead of a high density peak at the ionisation front, after $\sim$ 3 yrs the entire ionised region settles to a stable higher density profile. Figure \ref{fig:2temp} shows the final accretion profile follows closely the analytic solution for steady state two-temperature Bondi accretion \citep{Vandenbroucke2019}. As this higher density region builds up, the luminosity required to ionise it will have to increase until the steady state two-temperature accretion structure is reached. Figure \ref{fig:Qdevelopment} shows that the ionising luminosity required to maintain this structure increases in time to an almost constant value. If we stop calculating the luminosity, but keep it constant once the steady-state solution is reached, in time the simulation returns to the oscillatory behaviour described in section~3.2 due to numerical noise in the rad-hydro code.       

In conclusion we only reach a steady-state two-temperature accretion solution with careful fine-tuning of the ionising luminosity, and once the solution is reached it is not stable, evolving into a runaway instability to any fluctuation (including numerical noise). 

\subsection{Late time evolution}

As material accretes onto stars and they grow in mass, we expect their luminosity to grow correspondingly, meaning they can ionise increasingly larger volumes. As previously mentioned this increase in mass and luminosity happens over greater timescales than those we are considering, hence we simulate the evolution of the ionised region piecewise, keeping the mass and luminosity constant during each stage. We have so far only considered H{\sc ii} regions close enough to the star for the infall velocity to exceed $v_R$, where the ionisation front exhibits an R-type behaviour. In this phase we have found that the H{\sc ii} region is gravitationally trapped and oscillating. Once the relative velocity of the ionisation front decreases below $v_D$ it will transition to D-type. At this point gravity is no longer dominant and the ionisation front experiences a pressure driven expansion. For velocities between $v_R$ and $v_D$ there is no single front solution to the jump across the ionisation boundary. 

\begin{figure}
\includegraphics[width=\columnwidth]{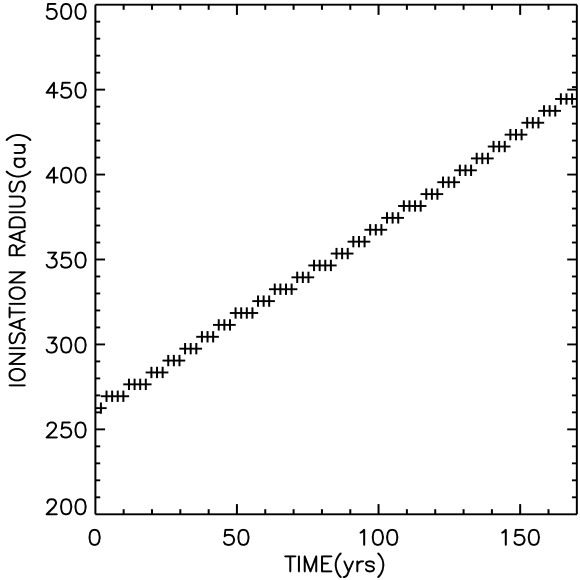}
	\caption{Ionisation front radius as a function of time during the late time simulation of an 17.87 $\rm{M}_{\odot}$ central star with constant ionising luminosity. Time zero years marks the moment radiation is switched on.}
	\label{fig:HIIexpand}
\end{figure}

\begin{figure*}
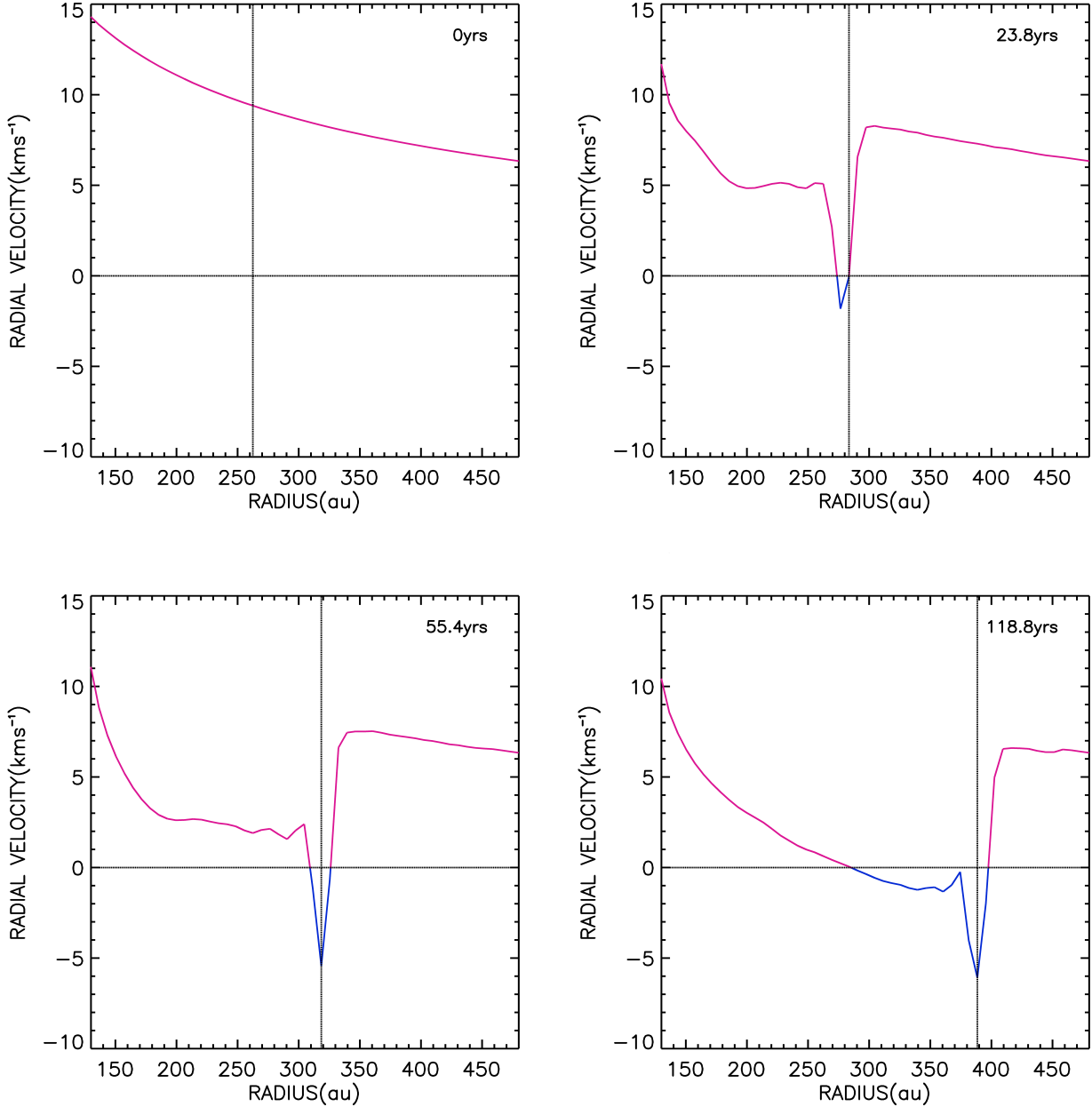

  \subfloat
  {\includegraphics[width=\columnwidth]{velocitystep1.pdf}}
  \subfloat
  {\includegraphics[width=\columnwidth]{velocitystep2.pdf}}
  
    \subfloat
  {\includegraphics[width=\columnwidth]{velocitystep3.pdf}}
  \subfloat
  {\includegraphics[width=\columnwidth]{velocitystep4.pdf}}
	\caption{The radially averaged velocity profile at four different times during the late time H{\sc ii} region expansion, a positive radial velocity corresponds to inward motion. Zero years marks the moment radiation is switched on. The simulation models an 17.87 $\rm{M}_{\odot}$ central star with a constant ionising luminosity large enough for the ionisation radius to exceed the ionised sonic point ($\sim 200$ AU). The vertical line represents the position of the ionisation front. The horizontal line marks zero velocity, the point across which the velocity changes direction. The profile is pink where the velocity points towards the central star, and blue where the velocity points away from the star.}
	\label{fig:velocityexpand}
\end{figure*}

\begin{figure*}
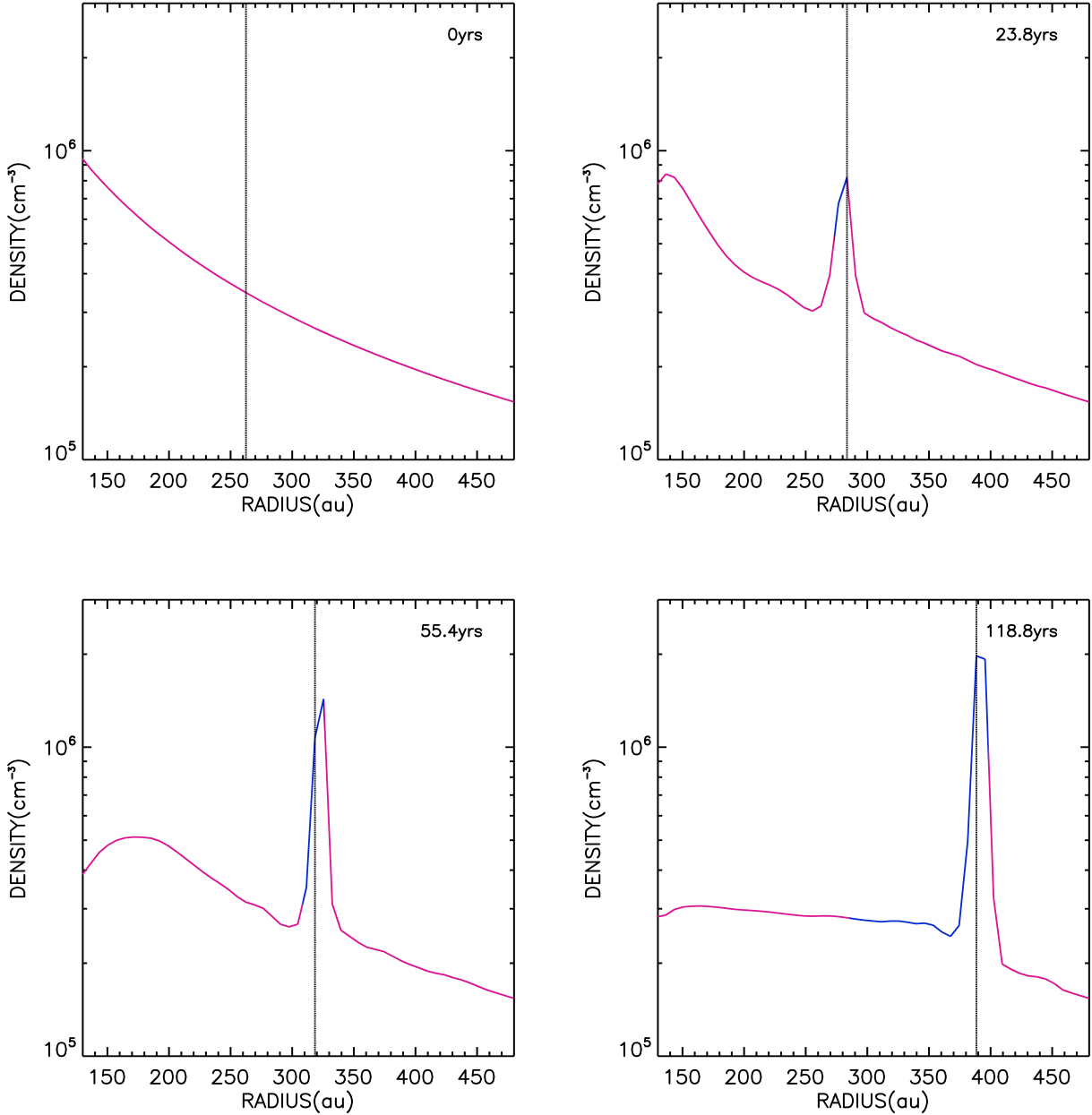

  \subfloat
  {\includegraphics[width=\columnwidth]{densitystep1.pdf}}
  \subfloat
  {\includegraphics[width=\columnwidth]{densitystep2.pdf}}
  
    \subfloat
  {\includegraphics[width=\columnwidth]{densitystep3.pdf}}
  \subfloat
  {\includegraphics[width=\columnwidth]{densitystep4.pdf}}
	\caption{The radially averaged density profile at four different times during the late time H{\sc ii} region expansion. Zero years marks the moment radiation is switched on. The simulation models an 17.87 $\rm{M}_{\odot}$ central star with a constant ionising luminosity large enough for the ionisation radius to exceed the ionised sonic point ($\sim 200$ AU). The vertical line represents the position of the ionisation front. The profile is pink where the velocity points towards the central source, and blue where the velocity points away from the source. Note the decreasing density in the inner H{\sc ii} region where material is accreting and hence draining the H{\sc ii} region and shutting off accretion.}
	\label{fig:densityexpand}
\end{figure*}

In order to investigate the evolution of the H{\sc ii} region past the R-critical radius we repeated the simulation described in section~3.2 for different luminosity values ionising beyond this radius. The size of the simulation box was increased to ${x,y,z} \in [-480,480]$~AU, which in turn increased the inner and outer mask radii to 124.8 AU and 456.0 AU. We found the H{\sc ii} region remains trapped and oscillating beyond the R-critical radius, until the ionised region extends beyond the ionised sonic point, at which point rapid expansion begins. The sonic point, where the infall velocity equals $c_i = 11.5\,\rm{km\,s}^{-1}$, lies approximately at 200 AU. The results presented in Figures \ref{fig:HIIexpand}-\ref{fig:densityexpand} are from a simulation using a luminosity of $Q = 2\times 10^{46}\,\rm{s}^{-1}$, where the initial ionisation radius reaches approximately 260 AU, exceeding the sonic point and hence resulting in rapid expansion. The luminosity is smaller than that previously needed to ionise out to $\sim$ 30 AU because we are only ionising material outside the inner mask, which in this simulation is further away from the star and thus at a much lower density. 

The simulation shows an expanding ionisation radius which is plotted as a function of time in Figure \ref{fig:HIIexpand}. The discrete jumps the ionisation front makes every 3-4 time steps is a result of the grid resolution. During the expansion phase the neutral material is moving towards the central star due to gravity just as it would if there was no radiation, until it reaches the expanding dense shell. The shell, along with most of the ionised material it accumulates, moves radially outwards, but close to the star ionised material still accretes. Figures \ref{fig:velocityexpand} and \ref{fig:densityexpand} show the directions of motion of the material at four different time steps, with red and blue indicating inflow and outflow respectively. The material inside the H{\sc ii} region is not replenished, since no material crosses the ionisation front, meaning the region is eventually drained.

The transition point from trapped to expanding can be used to estimate the time scale of the UCH{\sc ii} region phase. The \cite{Wood1989} survey indicates 10\% of all O and B stars are surrounded by an UCH{\sc ii} region. This suggests the lifetime of UCH{\sc ii} regions must exceed $10^5$ yrs, which is longer than the time it would take an UCH{\sc ii} region to evolve to a diffuse H{\sc ii} region through pressure driven expansion \citep{Wood1989}. During the time it takes to accrete enough material onto our central star to reach a high enough mass corresponding to a high enough luminosity to ionise past the sonic point, the H{\sc ii} region will remain trapped. Including this period before the expansion phase begins increases the total lifetime of UCH{\sc ii} regions.

\section{Conclusions}

We have presented the first look at the steady-state models of H{\sc ii} region evolution within spherically symmetric Bondi accretion flows by \cite{Keto2002b,Keto2003} and \cite{Mestel1954} in a radiation hydrodynamic framework. Our numerical radiation hydrodynamic simulations reproduce the main features of the analytic steady-state analyses, namely gravitationally trapped H{\sc ii} regions at early times and pressure-driven expansion when the ionising luminosity increases to produce an H{\sc ii} region extending beyond the critical radius at approximately the ionised sonic point. 

The main difference between the analytical work and our simulations is the stability of the trapped H{\sc ii} region. Any increase in density within the ionised region, whether due to jump conditions across the ionisation front or noise in the simulation, initiates a runaway instability of the H{\sc ii} region. This instability is expected to occur only in idealised spherically symmetric geometry.

Regardless of the H{\sc ii} region oscillation, our results are in broad agreement with the analyses by \cite{Keto2002b,Keto2003} and \cite{Mestel1954}. The implication of this picture on massive star formation is that accretion of material onto forming stars can continue even after the star starts emitting ionising radiation, see also \cite{Kuiper2018}. This is because the H{\sc ii} region remains gravitationally trapped, and all velocities are inflowing, until the ionising luminosity grows (due to mass accretion) and the radius of ionisation passes the critical radius where pressure-driven expansion begins. At this stage the accretion rate onto the star decreases to zero and the star reaches its maximum accreted mass. This could change estimates of the accretion rate during the UCH{\sc ii} phase of formation of high-mass stars as the accretion period can last for longer. Furthermore, gravitational trapping could also help explain the surprisingly long lifetimes of UCH{\sc ii} regions. 

The next paper in this series will investigate the discovered instability analytically as well as with a 1D code capable of much higher resolution than our idealised 3D simulations. The simulations presented in this paper produce spherically symmetric ionised accretion flows in an isolated system, no rotation or magnetic fields are introduced. This simplified approach is necessary for a direct comparison to the analytical model presented by \cite{Keto2002b,Keto2003} and \cite{Mestel1954}, however it is not a realistic treatment of the evolution of H{\sc ii} regions. Future work will explore radiation hydrodynamic simulations of rotationally flattened ionised accretion structures comprising equatorial inflow and simultaneous bipolar outflows \citep{Keto2007} and the production of synthetic continuum and line intensity maps to compare with observations.

\section*{Acknowledgements}

We thank the anonymous referee for helpful comments. KL acknowledges financial support from the Carnegie Trust. KW, DFG and BV acknowledge support from STFC grant ST/M001296/1. D.F.G. thanks the Brazilian agencies CNPq
(no. 311128/2017-3) and FAPESP (no.  2013/10559-5) for financial support. NS thanks CAPES for the financial support. 
\pagebreak



\bibliographystyle{mnras}
\bibliography{refs.bib}



\bsp	
\label{lastpage}
\end{document}